 \documentstyle[prl,aps,multicol,epsfig]{revtex}
 \begin{document}
 \draft

\title{Strong magnetic field in a protoneutron star}
\author{Subrata Pal$^1$, Debades Bandyopadhyay$^1$, and 
Somenath Chakrabarty$^2$}
\address{$^1$Saha Institute of Nuclear Physics, 1/AF Bidhannagar, 
Calcutta- 700 064, India}
\address{$^2$Department of Physics, University of Kalyani, 
Kalyani 741235, India}
\address{and Inter-University Centre for Astronomy and Astrophysics,
Post Bag 4, Ganeshkhind, Pune 411007, India}
\maketitle

\begin{abstract}
We investigate the influence of a strong interior magnetic field on the
structure and composition of a protoneutron star allowing quark-hadron 
phase transition. In contrast to protoneutron stars with noninteracting 
quark phase, the stars with interacting quark phase have smaller maximum 
star masses than those of neutrino-free stars. The strong field makes the 
overall equation of state softer compared to the field-free case favoring 
the evolution of a protoneutron star more towards a neutron star than a 
black hole. 
\vspace{0.4cm}



\end{abstract}
\begin{multicols}{2}

The understanding of the final journey of a massive star,
after its fuel has been exhausted is a challenging problem \cite{Bet90}. 
The outcome of it may be a supernova and the residue either be a neutron star 
or a black hole. The unique feature about supernova problem is that it involves 
all the forces of nature $-$ strong, weak, electromagnetic and gravity. A 
massive star in its late stage of evolution undergoes gravitational core 
collapse as the core exceeds the Chandrasekhar mass. The subsequent core 
bounce occurs when the core density reaches nuclear matter value and above. 
A few milliseconds after the core bounce, the hot and lepton rich 
(neutrino-trapped) core settles into hydrostatic equilibrium \cite{Bur}. 
The evolution of this protoneutron star into a neutron star or a black 
hole completes within a few tens of seconds.

Observations of pulsars predict large surface magnetic fields of $B_m \sim 
10^{14}$ G \cite{Chan}. The interior magnetic fields are a few orders of 
magnitude larger than the surface fields. In fact, the virial theorem 
\cite{Chand} predicts large interior field of $\sim 10^{18}$ G or more 
\cite{Cha}. One of the plausible explanations 
for such a large interior field is that the weak
field of a progenitor is amplified because of flux conservation during
the gravitational core collapse. The highly conducting core results in 
large ohmic diffusion time so that the field is frozen in the core, and
consequently not manifested at the surface \cite{Bay1,Mus}. 
The energy of a charged particle changes significantly in the quantum 
limit if the magnetic field is comparable to or above a critical value 
$B_m^{(c)}$, and the quantum effects are most pronounced when the 
particle moves in the lowest Landau level \cite{Cha}. In addition to 
this large magnetic field embedded in the dense core of the 
(proto)neutron star, a transition from nuclear matter to a stable 
quark matter is also possible \cite{Gle,Pra,Dra,Ban}.

In this letter we investigate the effects of trapped neutrinos on the
composition and structure of a protoneutron star in presence of a strong
interior magnetic field and also allowing a hadron to quark phase
transition in the star's interior. The matter inside the 
protoneutron star is highly degenerate and the chemical potential of 
its constituents are a few hundreds of MeV. On the other hand, the central
temperature of the star is a few tens of MeV. Therefore, neglecting
finite temperature effect will have little influence on the gross 
properties of the star. However, the compositional changes caused
by trapped neutrinos, which are primarily of the electron type $\nu_e$, 
may induce relatively larger changes in the maximum masses \cite{Pra}.  

We describe the equilibrium conditions of the pure quark and baryonic
matter and their mixed phase coexisting in a uniform background of 
electrons ($e$) and electron neutrinos ($\nu_e$) for neutrino-trapped 
matter in a uniform magnetic field $B_m$ along $z$ axis. The 
neutrino-trapped pure quark phase consisting of $u$, $d$ and $s$ quarks 
interacting through one-gluon exchange in local charge neutral 
and $\beta$-equilibrium conditions is described by the bag model \cite{Far}. 
The interaction energy density due to one-gluon exchange term to order $g^2$ 
for each flavor with $B_m\neq 0$ in the zeroth Landau level, $\eta=0$, 
is given by \cite{Ban}
\begin{equation}
 {\cal E}^{f;\eta=0}_I = {q_f B_m\over 8\pi^2} \int^{+p_F^f}_{-p_F^f} dp_z
\left[ U^F_0 + {p_{vz}\over \sqrt{p^2_{vz} + m_f^{* 2}}} U^F_V \right], 
\end{equation}
where $q_f$, $m^*_f$ and $p_F^f$ are the charge, effective mass and 
Fermi momentum of quark of $f$th flavor, and 
$p_{vz}=p_z[1+U^F_V/p_z]$. For the expressions of the Fock
contributions, $U^F_0$ and $U^F_V$, to the single particle energy, we 
refer to Ref. \cite{Ban}.
The QCD coupling constant is defined by $\alpha_c = g^2/4\pi$. The general 
expression (for all Landau levels $\eta$) for the total kinetic energy 
of the quark phase in a magnetic field is
\begin{eqnarray}
{\cal E}^\eta_K &=& \sum_{f=u,d,s} {d_fq_fB_m\over 4\pi^2}
\sum^{\eta_{\rm max}^{(f)}}_{\eta=0} g_{\eta} 
\Phi\left(\mu_f^*,m^*_{f,\eta}\right) \nonumber\\
&+& {eB_m\over 4\pi^2} \sum^{\eta_{\rm max}^{(e)}}_{\eta=0} g_{\eta}
\Phi\left(\mu_e,m_{e,\eta}\right) + {\mu_{\nu_e}^4\over 8\pi^2} ~,
\end{eqnarray}
where $\Phi(x,y) = x{\cal O}_{i,\eta}^{1/2} + y^2 \; {\rm ln} \left\{ 
\left( x + {\cal O}_{i,\eta}^{1/2}\right) \Big/y \right\}$  with $i\equiv f$ or $e$.
The notation in Eq. (2) is same as that in Ref. \cite{Ban}, with the first 
term corresponds to those for quarks with $d_f=3$, the second term is that
for electrons, and the third term corresponds to that for the neutrinos. 
The total energy density of the pure quark phase is then 
${\cal E}^q = {\cal E}^{\eta}_K + \sum_f {\cal E}_I^{f;\eta=0} + {\cal E}_m + B$, 
where ${\cal E}_m = B_m^2/(8\pi)$ is the magnetic field energy density and
$B$ is the bag constant. The pressure follows from the relation
$P^q = \sum_f \mu_f n_f + \sum_l \mu_l n_l - {\cal E}^q$, where $\mu_f$ denotes  
the quark chemical potential,  
$n_f = \left( d_fq_fB_m/2\pi^2 \right) \sum_{\eta=0}^{\eta_{\rm max}^{(f)}} 
g_{\eta} \left( \mu_f^{* 2} - m_{f,\eta}^{* 2} \right)^{1/2}$ is the quark 
density, and $l\equiv (e,\nu_e)$. The electron density is 
$n_e = \left( eB_m/2\pi^2 \right) 
\sum_{\eta=0}^{\eta_{\rm max}^{(e)}} g_{\eta} \left( \mu_e^2 
- m_{e,\eta}^2 \right)^{1/2}$, and $\mu_e$ and $\mu_{\nu_e}$ are the chemical 
potentials for electrons and electron neutrinos.
The charge neutrality condition, 
$Q^q = \sum_f q_f n_f - n_e = 0$, and the $\beta$-equilibrium conditions,
$\mu_d = \mu_u + \mu_e - \mu_{\nu_e} = \mu_s$, can be solved self-consistently 
together with the effective masses at a fixed baryon number density 
$n_b^q = (n_u+n_d+n_s)/3$ to obtain the equation of state (EOS) for the 
deconfined phase. For the ease of numerical computation we, however, 
add here the one-gluon exchange term perturbatively to energy density 
and pressure.

To describe the neutrino-trapped pure hadronic matter consisting of 
neutrons ($n$), protons ($p$), electrons ($e$) and $\nu_e$, we employ 
the linear $\sigma$-$\omega$-$\rho$ model of Ref. \cite{Zim} in the 
relativistic Hartree approach. The Fock contribution to the hadron phase 
is quite small \cite{Ser}, and therefore neglected. The EOS for this phase is 
obtained by solving self-consistently the effective mass in conjunction with 
the charge neutrality and $\beta$-equilibrium conditions, $Q^h = n_p - n_e = 0$ 
and $\mu_n = \mu_p + \mu_e - \mu_{\nu_e}$ at a fixed baryon number density 
$n_b^h$. Here $n_i$ and $\mu_i$ denote the number density and chemical potential;
the subscript $i$ refers to $n$, $p$, $e$ and $\nu_e$. The total energy density 
${\cal E}^h$ (given in Ref. \cite{Ban}) and pressure $P^h$ in this phase are 
related by $P^h = \sum_i n_i\mu_i - {\cal E}^h$.

The mixed phase of hadrons and quarks comprising of two conserved charges,
baryon number and electric charge is described following Glendenning \cite{Gle}.
The conditions of global charge neutrality and baryon number conservation 
are imposed through the relations $\chi Q^h + (1-\chi)Q^q = 0$ and 
$n_b = \chi n_b^h + (1-\chi)n_b^q$, where $\chi$ represents the fractional
volume occupied by the hadron phase. Furthermore, the mixed phase 
satisfies the Gibbs' phase rules: $\mu_p = 2\mu_u + \mu_d$ and $P^h = P^q$.
The total energy density is ${\cal E} = \chi{\cal E}^h + (1-\chi){\cal E}^q$. 
The neutrino-free matter relations can be obtained \cite{Ban}
by putting $\mu_{\nu_e} = 0$ in the above expressions.

In the present calculation the values of the dimensionless coupling constants 
for $\sigma$, $\omega$ and $\rho$ mesons determined by reproducing the nuclear 
matter properties at a saturation density of $n_0 = 0.16$ ${\rm fm}^{-3}$ are 
adopted from Ref. \cite{Gle}. The current masses of $u$ and $d$ quarks 
are taken as $m_u = m_d = 5$ MeV and $m_s = 150$ MeV, and the QCD coupling 
constant is $\alpha_c = 0.2$. Because of trapping the numbers of leptons
per baryon, $Y_{Le} = Y_e + Y_{\nu_e}$, are conserved on dynamical time scale. 
Gravitational collapse calculations of massive stars indicate that, at the
onset of trapping, $Y_{Le} \simeq 0.4$. We consider the bag constant 
$B=250$ MeV ${\rm fm}^{-3}$ which corresponds to the lower limit dictated 
by the requirement that, at low density, hadronic matter is the preferred phase.
The magnetic field, like the baryon density, increases from the surface
to the center of the star, and consequently the variation of the magnetic 
field $B_m$ with density $n_b$ is parametrized by the form \cite{Ban}
\begin{equation} 
B_m(n_b/n_0) = B_m^{\rm surf} + B_0 \left[
1 - \exp\left\{ -\beta (n_b/n_0)^{\gamma} \right\} \right] ,
\end{equation} 
where the parameters are chosen to be $\beta = 10^{-4}$ and $\gamma = 6$.
The maximum field prevailing at the center is taken as
$B_0=5\times 10^{18}$ G and the surface field is 
$B_m^{\rm surf} \simeq 10^{8}$ G. The number of Landau levels populated for 
a given species is determined by the field $B_m$ and baryon density \cite{Cha}. 
To the lowest order, the magnetic field energy density and magnetic pressure
have been treated perturbatively.

A traditional point of view has been that the magnetic field is frozen in 
the neutron star's interior,  
at least, during the age of the universe, i.e. $\sim 10^{11}$ years, 
because of high electrical conductivity in the core \cite{Bay1,Mus}. 
Only the crustal fields,  
$B_m \sim 10^{14}$ G, would diffuse to the surface and could be observed.
However, it was shown by Haensel et al. \cite{Hae} that the electrical
resistivity of the interiors of the star in the normal state (without 
superfluidity and superconductivity) can significantly increase. This
can lead to a dramatic decrease in the interior magnetic field decay time 
which is estimated to be 
$t \approx 2\times 10^{12} T_9^2 B_{12}^{-2} R_6^2 (n_b/n_0)^2$ years; in
the usual notation \cite{Hae}. Soon after the birth of a neutron star
when the temperature drops below the superconducting critical temperature
$T_c \approx 10^8 - 10^{10}$ (which is model dependent), it is commonly 
thought that superfluid protons form a type II superconductor in the 
outer core within the density range $0.7n_0 < n_b < 2n_0$,
and the estimated lower and upper critical magnetic fields are 
respectively $H_{c1} \sim 10^{15}$ G and $H_{c2} \sim 3\times 10^{16}$ G 
\cite{Bay2}; we have also obtained these values in the present model. 
Of course with increasing baryon densities, $n_b>2n_0$, the attractive
interaction between protons in the $^1S_0$ state (which gives rise to 
proton superconductivity) is diluted, this may result in the possible
transition to a type I superconductor; the critical field for which is,  
however, much smaller $H_c \sim 10^{14}$ G \cite{Sed}.

In our calculation, the magnetic field of $10^{18}$ G occurring at 
densities $n_b \approx 3n_0$ in the inner core (see Eq. (3)) is thus 
strong enough to suppress superconductivity. Therefore, for a neutron 
star of typical radius $R=10$ km and temperature $T=10^9$ K, the 
characteristic diffusion time for such field, using the above estimate 
of Ref. \cite{Hae}, is then $t \sim 20$ years, and thus should
have been observed at the surface of the neutron star after its birth. 
However, with our choice of the variation of the magnetic field 
with density (see Eq. (3)), at the {\it bulk of the outer core} 
($n_b \approx (0.7- 2)n_0$) the field is only $10^{14} - 10^{16}$ G. 
Therefore for $T<T_c$, the outer core could form a superconducting region. 
Thus soon after
the formation of the neutron star, the superconducting outer core region 
can, in principle, trap the much stronger field at the inner core and 
center (in the normal state) preventing its decay to the surface. 
However, the situation becomes much more complex in a superconducting state. 
The homogeneous magnetic field then splits into an ensemble of fluxoids 
\cite{Bay2} which contain a strong magnetic field, embedded in the 
field-free superconducting medium for a type II superconductor. It was also
shown \cite{Mus85} that such fields can be expelled out of the 
superconducting region to the subcrustal region in the first $\sim 10^4$ 
years, and subsequently they decay to the surface.  
In any case in what follows, for the present study of the evolution of a 
protoneutron star, within the first few tens of seconds when the star is hot, 
into a neutron star or a black hole in presence of a strong interior uniform 
magnetic field, we assume that the large fields are frozen in the core and 
therefore would not decay to the surface.

With the given parameter set of Eq. (3), we show in Table I, the mixed phase 
boundaries for the protoneutron star ($Y_{Le} = 0.4$) and neutrino-free star 
($Y_{\nu_e} =0$) for $B_m=0$ and $B_0=5\times 10^{18}$ G with interacting
quark phase (IQP). To examine the measure of the importance of the one-gluon 
exchange contribution, the corresponding results for stars with noninteracting 
quark phase (NQP) are given in parentheses. The onset of transition is
at density $n_1 = u_1n_0$, and the pure quark phase begins at $n_2=u_2n_0$, 
where $n_0=0.16$ ${\rm fm}^{-3}$ is the nuclear matter saturation density. 
In contrast to NQP matter, with the inclusion of interaction, the EOS for the
quark phase becomes softer in the corresponding IQP matter and the transition
to a mixed phase is delayed. At high density, as expected, 
the EOS for the quark phase is more softer resulting in a larger shift in 
$u_2$ to higher density compared to that in $u_1$. As a consequence, the 
extent of the mixed phase is increased in IQP than the corresponding NQP matter.
For both NQP and IQP matters, neutrino trapping delays the onset of the 
transition to higher baryon densities and also reduces the extent of the 
mixed phase in contrast to neutrino-free matter for both cases with and 
without magnetic field. 

The maximum masses of the stars $M_{\rm max}$, and their central densities 
$n_c=u_cn_0$ obtained by solving the Tolman-Oppenheimer-Volkoff equation 
are also given in Table I for different cases studied. For all cases 
considered, since the central densities $u_c$ are less than $u_2$ and fall 
within the mixed phase, the presence of a pure quark phase is precluded. 
With only nucleons
and leptons, without any quark, the maximum masses and central densities 
of neutrino-trapped(free) stars are respectively $M_{\rm max}=1.699
(1.778)M_\odot$ and $u_c=10.400(9.801)$ for $B_m=0$ while those for 
$B_0=5\times 10^{18}$ G are $M_{\rm max}=1.545(1.650)M_\odot$ and 
$u_c=11.075(9.052)$. Therefore, for pure hadronic matter, neutrino
trapping generally reduces the maximum mass from that of the neutrino-free 
star, with and also without magnetic field. This is caused by the conversion 
of protons to neutrons, as the trapped neutrino leave, which increases
the pressure more than it decreases by the loss of neutrinos.
With the introduction of the quarks, which soften the EOS, the
scenario becomes quite interesting. For the NQP, the maximum mass for the
neutrino-trapped star is found to be larger than that of the neutrino-free
star. This reversal in behavior from the pure hadronic case, was in fact, 
first noted by Prakash et al. \cite{Pra}, and an explanation 
sought was that the delayed appearance of the quarks in neutrino-trapped 
star resulted in a stiffer overall EOS. However, in contrast to NQP matter, 
we find for the IQP, the trend of the maximum mass is similar
to those for hadronic stars, i.e., after deleptonization the maximum mass
of star increases. In order to explain the behavior of $M_{\rm max}$, we 
consider the fraction of the gravitational mass originating from the pure
hadronic part, $M_{\rm had}/M_{\rm max}$, of stars with quark-hadron
phase transition. We consider first stars in {\it absence of magnetic
field}. The quarks here soften the EOS, thus larger contribution from
quarks (i.e. small $M_{\rm had}/M_{\rm max}$) would lead to smaller 
$M_{\rm max}$. As evident from Table I, for stars with NQP, a much larger 
\end{multicols}

\begin{table}

\caption{ The phase boundaries, $u_1$ and $u_2$, and central densities 
$u_c$ for neutrino-trapped ($Y_{Le}=0.4$) and neutrino-free 
($Y_{\nu_e}=0$) stars with maximum masses $M_{\rm max}/M_{\odot}$ 
with and without magnetic field that undergo a quark-hadron phase 
transition with interacting quark phase. The fraction of mass originating 
from the pure hadron phase is $M_{\rm had}/M_{\rm max}$. The corresponding 
quantities with noninteracting quark phase are shown in parentheses. 
The variation of magnetic field with density $n_b$ is given by Eq. (3) with 
$\beta = 10^{-4}$, $\gamma =6$ for $B_0=5\times 10^{18}$ G. Calculations are 
performed for a mean field model of baryons and a bag model of quarks with 
bag constant of $B=250$ MeV ${\rm fm}^{-3}$. The nuclear matter saturation 
density $n_0$ is 0.16 ${\rm fm}^{-3}$. }

\begin{tabular}{c|cccccc} 

\hfil& $B_m$ (G)& $u_1=n_1/n_0$& $u_2=n_2/n_0$& $u_c=n_c/n_0$& 
$M_{\rm max}/M_{\odot}$& $M_{\rm had}/M_{\rm max}$ \\ \hline
$Y_{Le}=0.4$& 0& 8.167(6.541)& 28.067(19.390)& 11.442(11.102)& 
1.677(1.664)& 0.816(0.698) \\
$Y_{\nu_e}=0$& 0& 4.107(3.329)& 25.941(17.681)& 7.152(7.754)& 
1.707(1.610)& 0.748(0.584) \\ \hline
$Y_{Le}=0.4$& $10^8-5\times 10^{18}$& 8.091(6.503)& 27.897(19.216)& 
12.003(11.060)& 1.540(1.531)& 0.773(0.686) \\
$Y_{\nu_e}=0$& $10^8-5\times 10^{18}$& 4.110(3.322)& 25.754(17.530)& 
6.202(6.304)& 1.553(1.487)& 0.859(0.718) 

\end{tabular}
\end{table}

\begin{multicols}{2}

\noindent fraction of mass originates from the softer quark-hadron mixed 
phase in neutrino-free stars than neutrino-trapped matter. Consequently,
$M_{\rm max}=1.610M_\odot$ for neutrino-free star with NQP is smaller 
than $1.664M_\odot$ mass for protoneutron star. On the other hand,
for stars with IQP (with $B_m=0$), the pure hadronic part mostly governs
the EOS (i.e. large $M_{\rm had}/M_{\rm max}$) for both neutrino-free
and trapped matter. As a result the maximum mass here follows the trend 
of pure hadronic stars with $B_m=0$, i.e. $M_{\rm max}$ for 
$Y_{Le}=0.4$ is smaller than that for $Y_{\nu_e}=0$. This also accounts 
for the fact that the masses of stars with IQP are larger than the masses
of stars with the corresponding NQP matter.

In {\it presence of magnetic field}, the hadronic EOS becomes much
softer compared to the stiffening caused in the quark EOS \cite{Ban}. 
Therefore, the {\it overall} EOS for the quark-hadron star with 
$B_m\neq 0$ turns out to be softer leading to smaller mass than that 
for the field-free star. Thus for $B_m\neq 0$, smaller the contribution 
to the EOS from stiff quark part of the mixed phase, smaller 
would be the mass. This, in fact, is revealed by stars with NQP for
$Y_{\nu_e}=0$ compared to $Y_{Le}=0.4$ case. The EOS of a star with IQP
for $B_m\neq 0$ is primarily governed by the pure hadronic part, their 
mass is thus reminiscent of that for the pure hadronic stars with 
$B_m\neq 0$. Therefore, the trend of the masses from neutrino-free to 
neutrino-trapped star with magnetic field are same as that of field-free
stars, both with interacting and noninteracting quark phase.

In Fig. 1 we depict a comparison of the composition of neutrino-free
matter (top panel) and neutrino-trapped matter (bottom panel) in a
magnetic field of $B_0=5\times 10^{18}$ G with IQP. In contrast to $B_m=0$ case
(not shown), the electron fraction, in particular, is enhanced in a magnetic 
field due to phase space modification \cite{Cha}. 
In the hadronic sector, 
the electron fraction increases because of neutrino trapping, which in 
turn increases the proton fraction due to charge neutrality condition 
and thereby reduces the neutron fraction due to baryon number conservation. 
On the other hand, $u$, $d$ and $s$ quarks themselves try to maintain 
charge neutrality, resulting in reduction of the electron fraction and 
consequently the neutrino fraction is enhanced in the mixed phase. Neutrino 
trapping also increases the $u$ quark abundance in comparison to the 
neutrino-free star.

The differences in the abundance of electron, in particular
within the inner core of different stars investigated, would be 
manifested in the total number of $\nu_e$ escaping during deleptonization.
The total number of $\nu_e$s emitted, $N_{\nu_e}$, from the inner core 
of mass $0.5M_\odot$ during deleptonization from the initial value of 
$Y_{Le}=0.4$ is obtained by integrating 
$n_{\nu_e} = n_b Y_{\nu_e} = n_b(Y_{Le}-Y_e)$ 
over this region of the neutrino-free star. For pure hadronic star, 
$N_{\nu_e}$ is found to be $9.416\times 10^{55}$ at $B_m=0$, while
for $B_0=5\times 10^{18}$ G it is $9.135\times 10^{55}$. On the other
hand, we obtain for stars with IQP(NQP), the value 
of $N_{\nu_e}$ to 
\par
{\centerline{
\epsfxsize=9.2cm
\epsfysize=10.8cm
\epsffile{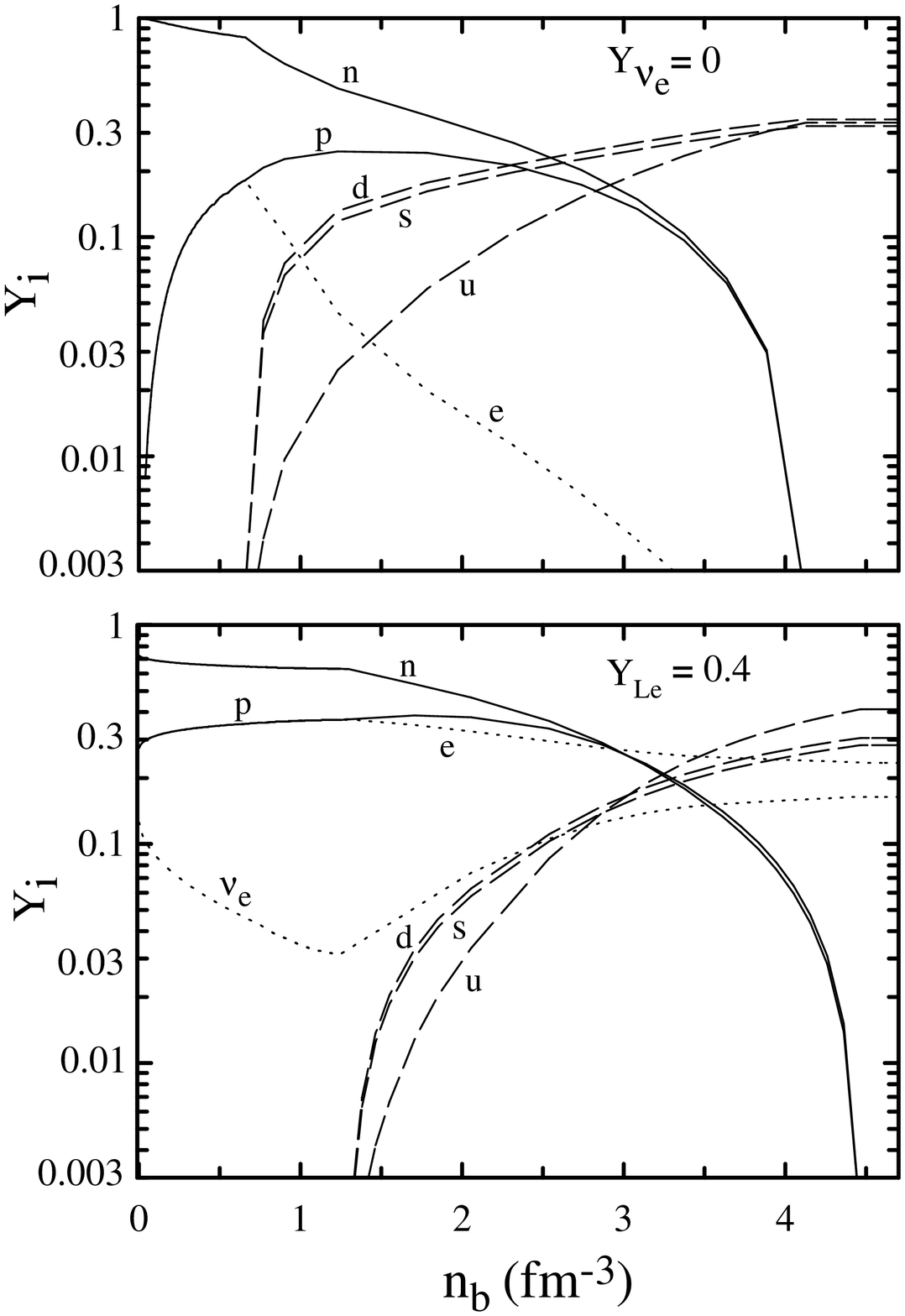}
}}
\vspace{-0.6cm}

\noindent{\small{Fig.$\:$1.$\!$ The abundances of neutrino-free (top panel) 
and neutrino-trapped
(bottom panel) star with interacting quark-hadron phase transition at 
a magnetic field $B_0=5\times 10^{18}~$G. }}
\vspace{0.65cm}

\noindent be $1.662\times 10^{56}(1.994\times 10^{56})$ for 
$B_m=0$, and for $B_0=5\times 10^{18}$G the corresponding values are 
$1.532\times 10^{56}(1.801\times 10^{56})$. The stars with quark-hadron 
phase transition emit more number of $\nu_e$s than that by the pure hadron 
stars. This is caused by the drop in the electron abundance in the mixed 
phase as quarks furnish net negative charge. The enhancement of 
electron fraction in magnetic field causes a decrease in $N_{\nu_e}$ from the
corresponding field-free cases. The different scenarios may thus be 
discernible from the difference in the neutrino numbers.

Delayed neutrino emission and possible black hole formation in the context 
of SN 1987A have been much debated issues in recent years. So 
far there is no observation of a pulsar in it. Moreover, the fading
away light curve leads one to think that it might have collapsed
into a low mass black hole. Assuming that SN 1987A has gone to a 
black hole, Bethe and Brown \cite{Bet95} estimated the gravitational mass
for the compact object in it to be $1.56M_{\odot}$ from Ni production.
They argued that any compact object having mass larger than this limit
would be a black hole. It has been counter argued \cite{Mus} that
the surface magnetic field of a nascent neutron star is weak. It may
take a few hundred years or more for the magnetic field trapped in the 
crust to reach the surface by ohmic diffusion which would then increase 
the surface field so that the pulsar could be accessible to observation.
In the present calculation, since the maximum masses of hadron stars 
increase when the trapped neutrinos leave, they will promptly collapse 
after core bounce into black holes once the mass of $1.56M_\odot$ is 
reached, presuming Bethe-Brown limit for the maximum neutron star mass. 
Similar situation arises for stars with interacting quark phase (see Table I). 
In contrast, the maximum masses of stars with noninteracting quark 
phase decrease after deleptonization. Therefore, in the event that 
the maximum masses of these neutrino-free stars reach $1.56M_\odot$, 
the stars will first explode, returning matter to the galaxy, and 
then collapse into low mass black holes \cite{Bet95}. We however find
in our calculation that the maximum star masses both for neutrino-trapped
and neutrino-free matter in presence of strong magnetic field are 
smaller (than $1.56M_\odot$) than those for the field-free stars.
Therefore, the presence of a strong interior field favors 
the evolution of a protoneutron star more towards a neutron star than 
a black hole even if its mass increases after the trapped neutrinos leave. 

Most of the dynamical supernova calculations \cite{Bet90,Bar,Arn} indicate 
that a successful prompt supernova explosion could be achieved by employing 
a soft EOS at densities larger than nuclear matter density $n_0$. The 
substantial softening of the EOS caused by the strong interior magnetic 
field (and also by quarks) may provide a viable prompt shock mechanism, 
and thereby merits consideration in dynamical simulations of explosions.

In summary, we have investigated the gross properties of a protoneutron
star in presence of a strong magnetic field in the core with a
quark-hadron phase transition. In contrast to the neutrino-trapped matter
with noninteracting quark phase, the protoneutron star with
interacting quark phase leads to smaller maximum star mass compared 
to neutrino-free case both with and without magnetic field.
Besides the presence of quarks, the softening of the overall EOS caused 
by magnetic field has significant bearing on the evolution of 
the protoneutron star to a neutron star	or a low mass black hole.

 

\end{multicols}
\end{document}